\documentclass{appolb}
\usepackage{graphicx}

\begin{document}
\title{Chiral spin symmetry and hot QCD
\thanks{Presented at Excited QCD 2024; Benasque, 14 - 20.01.2024 }%
}
\author{L. Ya. Glozman
\address{ Institute of Physics, University of Graz, A-8010 Graz, Austria }
\\[3mm]
}

\maketitle
\begin{abstract}
In this talk we overview main results indicating existence in QCD  of three qualitatively
different regimes  connected by smooth crossovers upon heating: a hadron gas, a stringy fluid and a quark-gluon plasma.
In the combined large $N_c$ and chiral limit these regimes likely become distinct phases separated
by phase transitions: a chiral restoration phase transition around $T_{ch} \sim 130$ MeV 
and a deconfinement phase transition around $T_d \sim 300$ MeV.
It should be an important  task  to verify this issue on the lattice.
We will introduce a chiral spin symmetry, which is a symmetry of the electric part of electrodynamics
and of QCD with light quarks.  It is realized
approximately in QCD above the chiral restoration crossover and disappears in the QGP
regime. The center symmetry of the pure glue action and the chiral spin symmetry
of the electric part of the QCD Lagrangian with light quarks are complementary to distinguish the confining regime and its disappearance.
We also address other  lattice evidences for stringy fluid: hadron resonances
extracted from the lattice correlators; breakdown of the thermal perturbation theory at  $T < ~ 600$ MeV and
fluctuations of conserved charges that point out the $N_c$ scaling above $T \sim 155$ MeV.
\end{abstract}
  
\section{Chiral spin symmetry}

Consider Maxwell equations that describe evolution of the electric and magnetic fields
in a given reference frame:

\begin{eqnarray}
div \vec E &=& 4\pi \rho \nonumber\\
rot \vec B - \frac{1}{c}\frac{\partial \vec E}{\partial t} &=& \frac{4\pi}{c}\vec j\nonumber\\
rot \vec E + \frac{1}{c}\frac{\partial \vec B}{\partial t} &=& 0 \nonumber \\
div \vec B &=& 0. 
\end{eqnarray} 
We define $\vec E$ and $\vec B$ in a given Lorentz frame in a gauge-invariant way through
their action on charge and current,
 
\begin{equation}
\vec F = q \vec E + q \frac{\vec v}{c} \times  \vec B.
\end{equation}
It is possible to measure directly $\vec F$ in electrodynamics but not possible in quantum chromodynamics.
Is there another method to distinguish $\vec E$ and $\vec B$? 

Consider charges to be massless particles with $s = 1/2$. They can be either
right- or left-handed:

\begin{equation}
\left(\begin{array}{c}
R\\
L
\end{array}\right).\nonumber
\end{equation}
Consider a $SU(2)_{CS}$ chiral spin transformation that mixes $R$ and $L$:

\begin{equation}
\left(\begin{array}{c}
R\\
L
\end{array}\right)\; \rightarrow
\left(\begin{array}{c}
R'\\
L'
\end{array}\right)=
\exp \left(i  \frac{\varepsilon^n \sigma^n}{2}\right) \left(\begin{array}{c}
R\\
L
\end{array}\right). 
\end{equation}\medskip
What happens with the  charge density $\rho$?
\begin{equation}
R'^\dagger R' + L'^\dagger L' = R^\dagger R + L^\dagger L,
\end{equation}
i.e.
\begin{equation}
\rho' = \rho.
\end{equation}
The charge density is invariant under the chiral spin transformation.
However, upon the chiral spin transformation the current density $\vec j$ and $\vec v$ change.
We  conclude that the interaction of a charge with the electric field
is invariant under the chiral spin transformation,
while interaction of a current with the magnetic field
is not. This means that the chiral spin symmetry allows one
to distinguish the electric and magnetic fields
in a given reference frame.
The electric part of the EM theory is more symmetric
than the magnetic part.

The chromoelectric field of QCD is defined via interaction with
the color charge 
\begin{equation}
 \vec F = Q^a \vec E^a;  ~~~~  Q^a = \int d^3x ~ 
q^\dagger(x) T^a q(x),~~~~a=1,...,8.
\end{equation}
The color charge, which is Lorentz-invariant, is  invariant under  $SU(2)_{CS}$:

\begin{equation}
~~~q \rightarrow  q^\prime = \exp \left(i  \frac{\varepsilon^n \Sigma^n}{2}\right) q,  \; ~~~~ \Sigma = \{\gamma_k,-i \gamma_5\gamma_k,\gamma_5\}.
\end{equation}
In a given Lorentz frame interaction of
quarks with the electric part of the gluonic field is  chiral spin invariant like in electrodynamics.
The  $SU(2)_{CS}$ symmetry includes $U(1)_A$ as a subgroup.

We can extend the $SU(2)_{CS}$ symmetry to
$SU(2N_F)$:
$
SU(2)_{CS} \times SU(N_F) \subset SU(2N_F)$ and 
$SU(N_F)_R \times SU(N_F)_L\times U(1)_A  \subset SU(2N_F)$.
The $SU(2N_F)$ is also a symmetry of the color charge and of the electric part of the QCD Lagrangian. I.e.,
the  color charge and electric part of the theory have a 
$SU(2N_F)$
symmetry that is larger than 
the chiral symmetry   of  QCD as a whole.
The fundamental
vector of $SU(2N_F)$ at $N_F=2$ is

\begin{equation}
\Psi^T =(u_R ~ u_L  ~ d_R  ~ d_L ). 
\end{equation}

Notice that the Dirac Lagrangian is not invariant under 
$SU(2)_{CS}$ and  $SU(2N_F)$ above: The chiral spin symmetry
and its flavor extension are explicitly broken by magnetic interactions and by the quark kinetic terms.

Since the confining interaction in QCD is associated with the emergence of
the color-electric flux tubes that connect static quarks we can consider
the chiral spin symmetry as a symmetry of the confining interaction with the
ultrarelativistic light quarks. The necessary and sufficient conditions
for emergence of the approximate chiral spin symmetry are: (i)  both chiral symmetries in QCD
must be at least approximately restored and (ii) the quark-electric interaction
must strongly dominate over the quark-magnetic interaction and over quark kinetic
terms. The latter condition means that the physics must be dominated by
a confining electric field. 

The discussed symmetries were introduced in Refs. \cite{G1,G2}. For  review
on symmetries and their implications for hadrons and for hot QCD see Ref. \cite{G3}.

\section{Emergence of the chiral spin symmetry above chiral
restoration crossover and its implications}
The most detailed information about QCD
is encoded in correlation functions  
\begin{equation}
C_\Gamma(t,x,y,z)=\langle O_\Gamma(t,x,y,z)\,O_\Gamma(0,\mathbf{0})^\dagger\rangle\;.
\end{equation}
They carry the full spectral information.

In Fig. 1 we show spatial correlators  of all possible $J=0,1$ isovector bilinears $O_\Gamma$ at different temperatures obtained within
$N_F=2$ QCD with chirally symmetric Dirac operator \cite{R2}.
\begin{figure}
  \centering
  \includegraphics[scale=0.25]{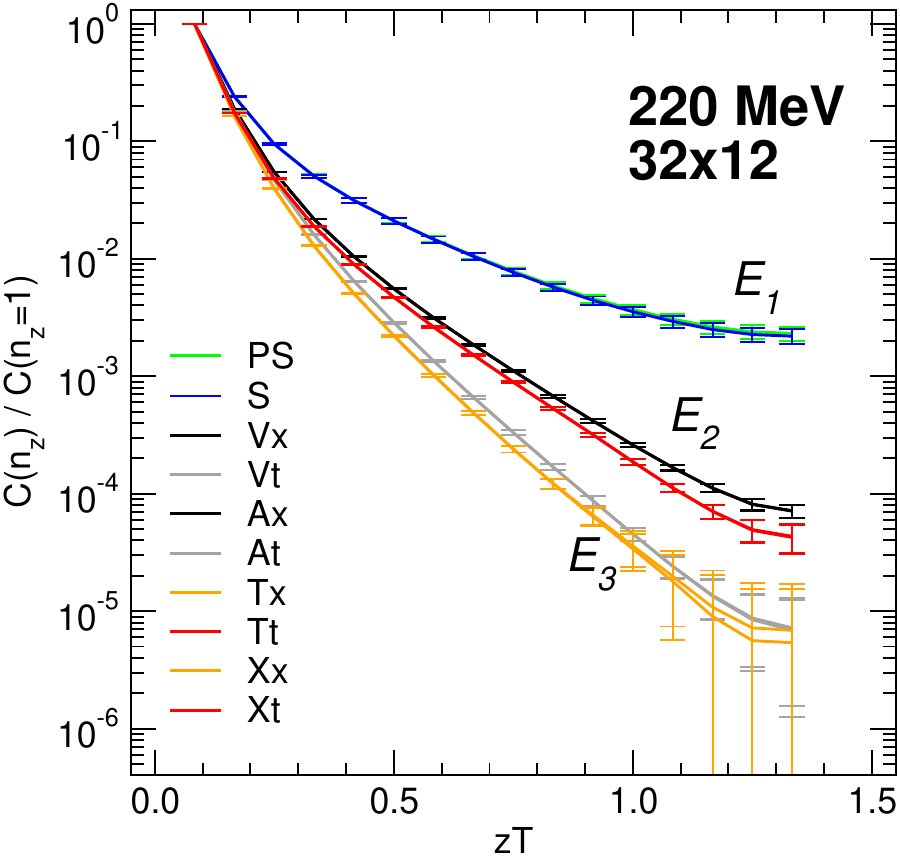} 
  \includegraphics[scale=0.25]{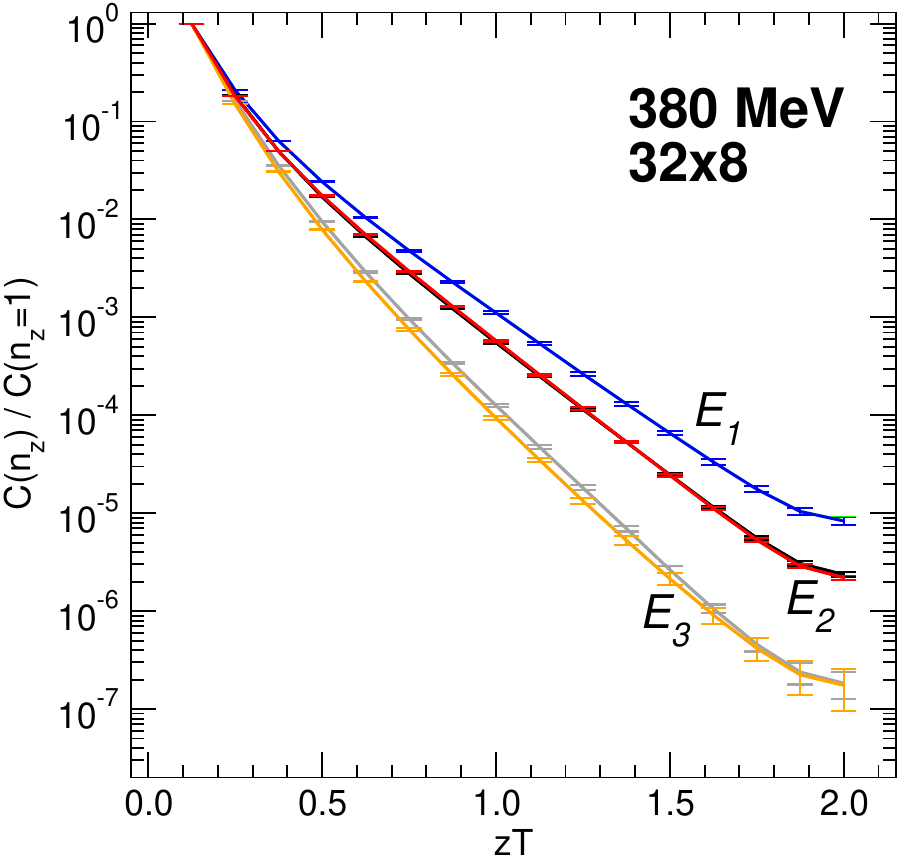} 
\caption{Spatial correlators of all possible $J=0,1$ bilinears. Source: From Ref. \cite{R2}.}
\end{figure}
We see a distinct multiplet structure of the correlators. The multiplet
$E1$ evidences  restored $U(1)_A$  symmetry (at least approximately); the multiplet  $E2$ demonstrates approximate chiral spin and $SU(4)$ symmetries; the multiplet $E3$ is consistent with
 chiral symmetry alone and with chiral spin (and $SU(4)$) symmetry and hence could be ignored. 
The chiral spin symmetry and its flavor extension 
 persist up to $T \sim
500$ MeV. At larger temperatures two distinct multiplets $E1$ and $E2$ disappear because
the QCD correlators approach correlators obtained with free quarks. Notice that the latter correlators
do not have the chiral spin symmetry because the Dirac Lagrangian is not CS-symmetric; for analytic and lattice
results for all possible "free correlators" see Ref. \cite{R2}. An apparent mergence
of the $E1$ and $E2$ multiplets at very high temperatures is due to the fact that at $T \rightarrow \infty$ all $J=0,1$ screening
masses from $E1$ and $E2$ approach their limiting value $2\pi T$, which is not related to any
dynamical symmetry.

In Fig. 2 we show temporal correlators for all possible
$J=0,1$ isovector bilinears \cite{R3}.
\begin{figure}
\centering  \includegraphics[width=0.3\linewidth]{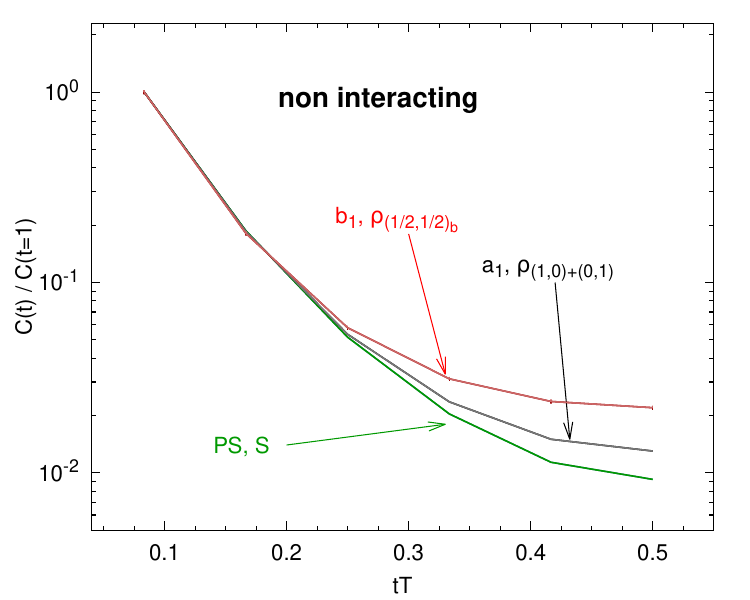}
\includegraphics[width=0.3\linewidth]{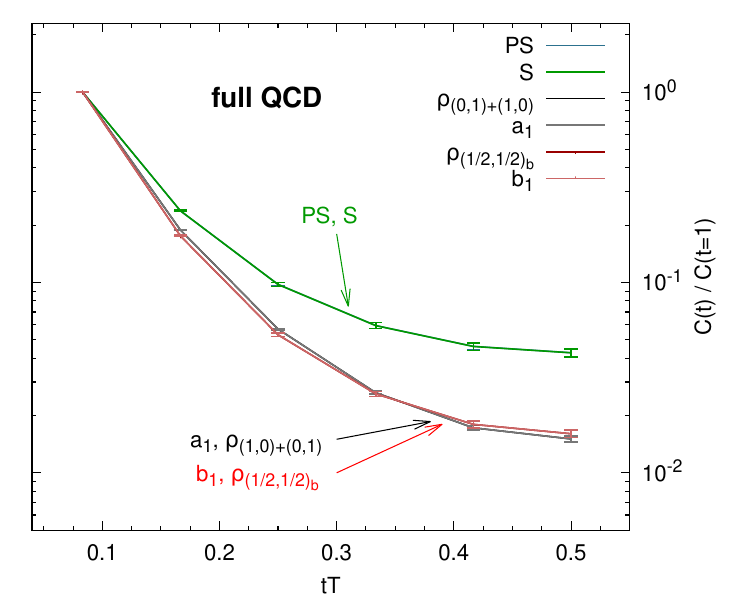}
  \label{tcorrs}
\caption{Temporal correlators for $12 \times 48^3$ lattices. Left:
correlators obtained with free quarks. Right: full QCD correlators 
at $T=220$ MeV. Source: From Ref. \cite{R3}}
\end{figure}
The full QCD correlators demonstrate
$U(1)_A$ and $SU(2)_L \times SU(2)_R$ multiplets as well as approximate
$SU(2)_{CS}$ and $SU(4)$ multiplets. In contrast, qualitatively different
correlators obtained with noninteracting quarks exhibit only
$U(1)_A$ and $SU(2)_L \times SU(2)_R$ multiplets.

We conclude that above the chiral restoration  crossover
QCD partition function is not only chiral symmetric, but is also
approximately  $SU(2)_{CS}$ and $SU(4)$ symmetric. This suggests that
QCD is still in the confining regime until roughly $500$ MeV where
confining chromoelectric field gets screened and $SU(2)_{CS}$ and $SU(4)$
symmetries smoothly disappear. Similar results have recently been obtained
in $2+1+1$ QCD \cite{Chiu}.

\section{Breakdown of thermal perturbation theory below $500 - 600$ MeV}

Lattice results for pseudoscalar and vector screening masses obtained at
very large temperatures  $T \sim$ 1 - 160 GeV \cite{brida}  are shown in Fig. 3,
left panel. 
\begin{figure}
  \centering
  \includegraphics[scale=0.275]{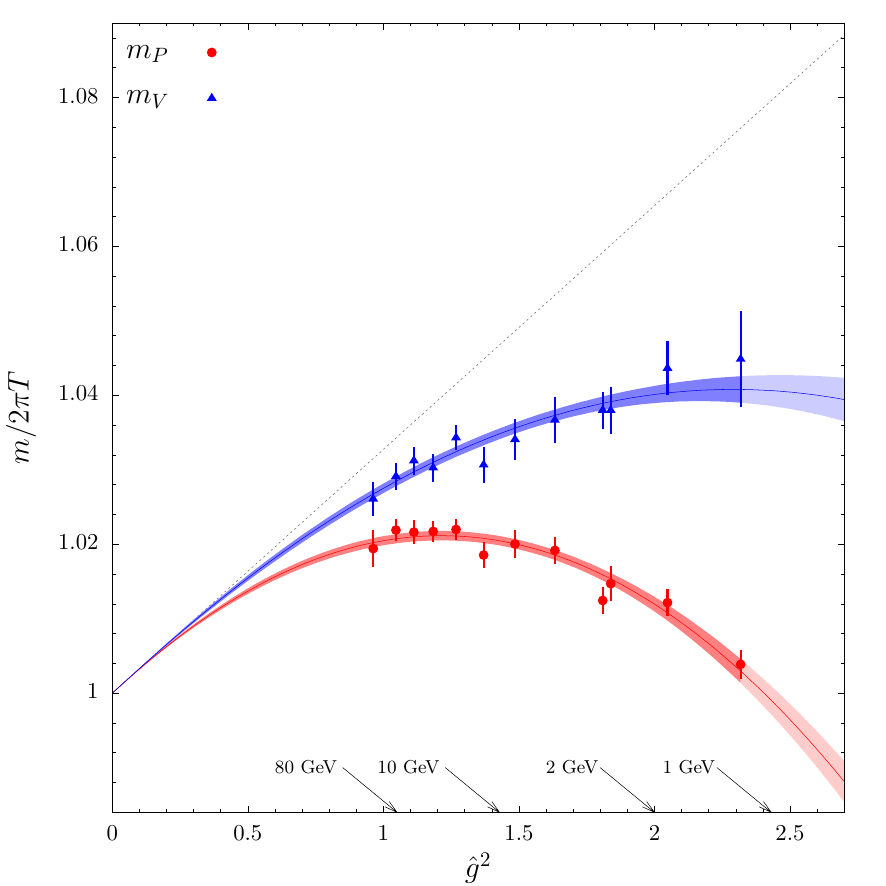}
   \includegraphics[scale=0.6]{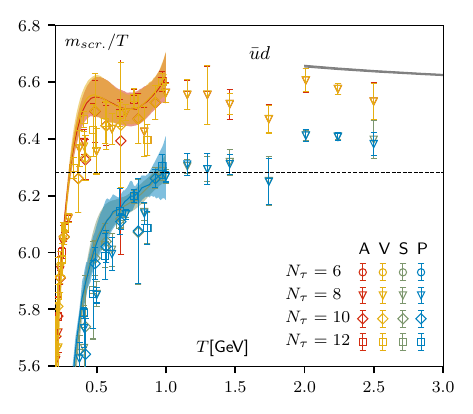} 
\caption{Left:Temperature dependence of the pseudoscalar and vector screening masses
at very large T. Source: From Ref. \cite{brida}. Right: The same but for lower temperatures.
Source: From Ref. \cite{Baz}}
\end{figure}
They can be parameterized \cite{brida} over two orders of magnitude by
\begin{eqnarray}
\frac{m_{PS}}{2\pi T}&=&1+p_2 \,\hat{g}^2(T) + p_3 \, \hat{g}^3(T)+p_4 \,\hat{g}^4(T)\;,\nonumber \\
\frac{m_{V}}{2\pi T}&=&\frac{m_{PS}}{2\pi T} + s_4\, \hat{g}^4(T)\;,
\nonumber
\label{eq:psv}
\end{eqnarray}
where $p_2$ is fixed by EQCD calculations and $p_3,p_4,s_4$ are numbers that are fitted
to lattice results. A perturbative description of screening masses suggests partonic degrees of freedom,
which is a signal of QGP.

Screening masses at lower temperatures, above chiral restoration crossover up to $ T \sim 2.5$ GeV
are shown in the right panel \cite{Baz}.
While the temperature dependence of screening masses above 500 - 600 MeV is flat and consistent
with those shown in the left panel,
at lower temperatures the screening masses demonstrate a steep increase. The temperature
dependence in the partonic (QGP) regime is only in coupling constants. Given that the coefficients
$p_2,p_3,p_4,s_4$ are fixed by the high T behavior of screening masses, the perturbative expansion
cannot explain screening masses below 500-600 MeV. We observe an apparent change of dynamics
at temperatures below 500-600 MeV. The behavior of meson screening masses from 12 different
channels provides an independent demonstration of the existence of the regime below 500-600
MeV where chiral symmetry is restored but the dynamics is inconsistent with the partonic description.
The discussed behavior of screening masses must be reflected in the equation of state. Indeed, a very
steep increase of $p/T^4$ with temperature in the same temperature interval is observed \cite{Baz2}. The analysis 
presented in this section was done in Ref. \cite{GPP}.

\section{Pion spectral function}
Direct evidence of the hadron-like degrees of freedom in stringy fluid should be observation of states in spectral functions. Using a finite T generalization
of the K\"allen-Lehmann spectral representation \cite{Bros} Refs. \cite{LP1,LP2} reconstructed
pion and kaon spectral functions from the spatial lattice correlators
above chiral restoration. The pion spectral function is shown
in Fig. 4 \cite{LP1}. The spectral function demonstrates
two distinct peaks corresponding to pion and its first radial excitations. These peaks get broader
with temperature and eventually melt above
500-600 MeV out. The so obtained spectral function can be controlled because it predicts the temporal correlators that can be compared
with lattice results. The comparison shows
a good agreement at large Euclidean  times. 
\begin{figure}
\centering
\includegraphics[scale=0.2]{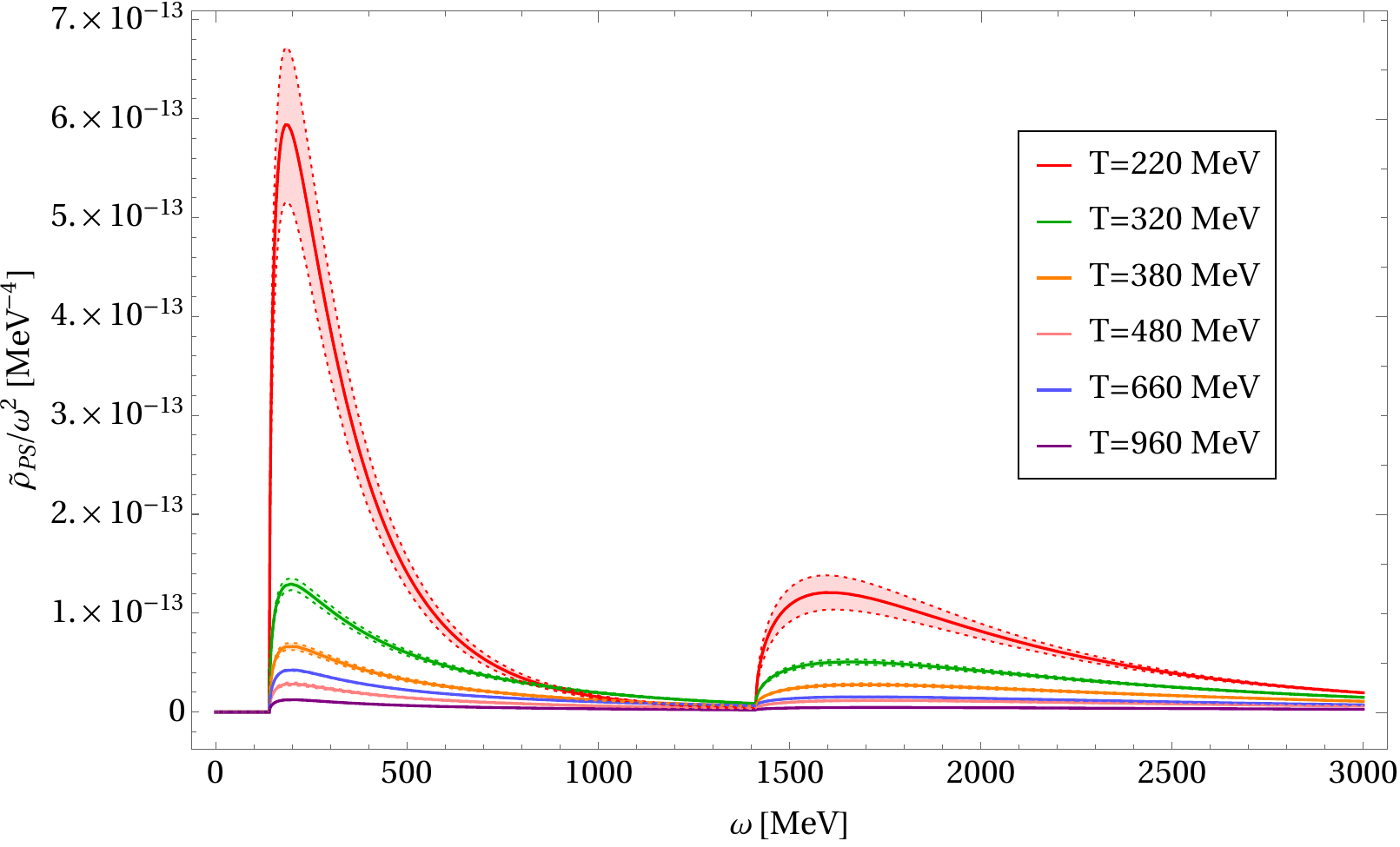}
\caption{Pion spectral function at different temperatures reconstructed
from spatial lattice correlators presented in Fig. 1. Source: From Ref. \cite{LP1}.}
\end{figure}

\section{Conserved charges and their fluctuations in hadron gas and at higher temperatures. Temperature evolution of the Polyakov loop.}

The hadron resonance gas model 
assumes a dilute system  of point-like structureless hadrons
that do not interact.  For $T < 155$ MeV the hadron resonance gas model  reproduces 
fluctuations of conserved charges calculated on the lattice, see e.g. \cite{Bel}.  At larger temperatures the lattice results
 radically deviate from the HRG model predictions.

Here \cite{CG2} we focus on charges associated with the net number of up, down and strange quarks:
\begin{equation}    
N_q \equiv \int d^3 x ~n_q(x) \; \;\;  {\rm with}
\; \; \;n_q(x) = \bar q(x) \gamma^0 q(x), \;\; \; q=u,d,s
\end{equation}

Each quark can be in one of the  $N_c$ color states. This means that the  conserved
flavor charges  $N_q$, scale as $N_c^1$.
The key point  is that the fluctuations of  quark number densities scale as $N_c^1$ above chiral crossover
and that in the stringy fluid phase at large $N_c$, they would be expected to differ from their vacuum and hadron gas values  (of order $N_c^0$) by $N_c$.  

In Fig. 5  we show typical results for fluctuations of quark numbers 
of $u,d$ quarks taken from Ref. \cite{Bel} and their comparison with the
hadron resonance gas model. We see
that the fluctuations of the  quark numbers deviate from
the HRG just at the chiral restoration temperature 155 MeV. 

\begin{figure}[h]
\centering
 \includegraphics[width=0.4 \textwidth]{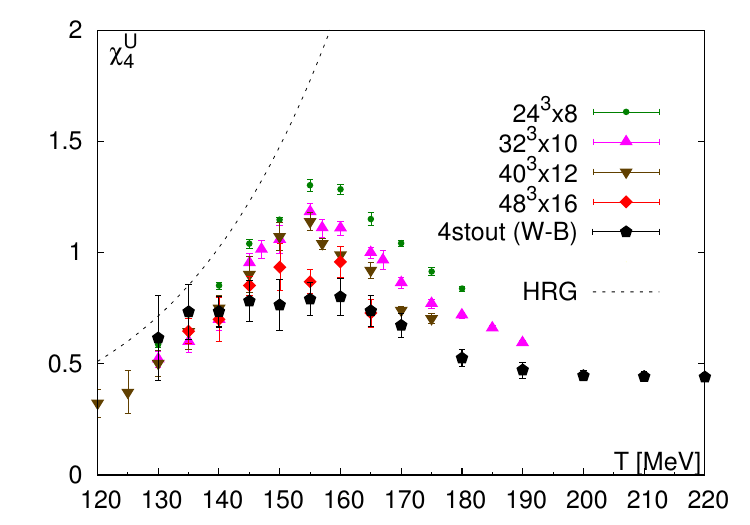}
 \includegraphics[width=0.34 \textwidth]{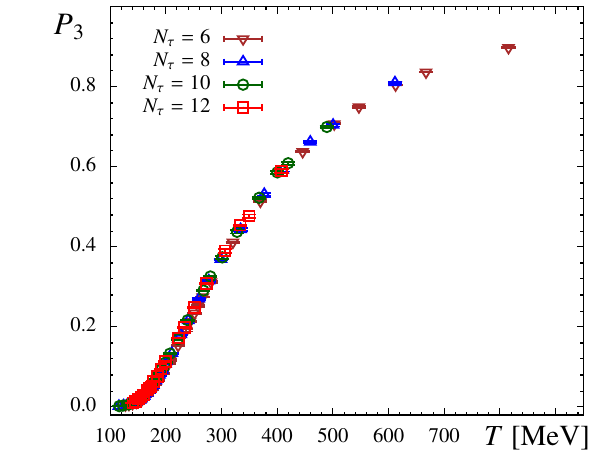}
 \caption{Left:Fluctuations of conserved net $u,d$ quark numbers in 2+1 QCD at physical quark masses. Source: From Ref. \cite{Bel}. Right: Temperature evolution of the properly renormalized Polyakov loop in 2+1 QCD at physical quark masses. Source: From Ref. \cite{PS}} 
    \label{fig2}
\end{figure}
Since the quark numbers scale as $N_c^1$ the above behavior of
fluctuations of conserved charges is consistent with the crossover from the hadron gas to the  stringy fluid regime,
where all main thermodynamical quantities scale as $N_c^1$ (it will be discussed in the next section). To demonstrate a transition to the
QGP regime one needs an observable that is sensitive to presence of $\sim N_c^2$ deconfined  gluons. A natural
observable of this kind is Polyakov loop.

 The expectation value of the Polyakov loop in a pure glue theory  is the order parameter for ${\cal Z}_{N_c}$ center symmetry and for confinement. In the center
 symmetric confining phase  the expectation value of the Polyakov loop identically vanishes while above the first order deconfinement phase transition at $T_d \sim  270 - 300 $ MeV the center symmetry gets spontaneously broken and the expectation value of the Polyakov loop jumps to 0.5-0.6.  An important issue is that the Polyakov loop in the 
deconfined phase, where it is not zero, is explicitly sensitive to $ N_c^2-1$
gluons.
 Confining
 properties of QCD with light quarks
 and of pure Yang-Mills theory are identical in the large $N_c$ limit. Then one expects  a similar
 deconfinement temperature in  QCD with light quarks at large $N_c$.

 In the real world $N_c=3$ in QCD with light quarks the center symmetry of the action is explicitly broken 
 by quark loops and 
the deconfinement first order phase transition 
 is replaced by a very smooth crossover.  The renormalized
Polyakov loop informs  us about the deconfinement crossover region.
The temperature evolution of the properly renormalized Polyakov loop, taken from Ref. \cite{PS}, is shown in Fig. 5. We observe that  indeed the deconfinement crossover is nothing else but a smeared broad  transition around $T_d \sim 300$ MeV.
Above the chiral restoration temperature around
$T_{ch} \sim 155$ MeV the Polyakov loop is very small which suggests that here QCD is
 in the confining regime. At the same time at these temperatures fluctuations of
 conserved charges demonstrate that the hadron gas picture does not work. Both these
 facts  are consistent with the crossover from the hadron gas to the stringy fluid. The
 Polyakov loop reaches the value around 0.5 at a temperature roughly $400-500$ MeV, in agreement
 with the temperature of smooth disappearance of the chiral spin symmetry.

\section{Large $N_c$ QCD phase diagram at $\mu_B=0$}
In previous sections we have demonstrated numerous
lattice evidences that upon increasing temperature there are three qualitatively different regimes in QCD that are connected by smooth crossovers: a hadron gas below $ T \sim
155$ MeV, an intermediate stringy fluid regime
at $ 155 < T < \sim 500$ MeV and a quark-gluon
plasma at higher temperatures. These regimes reflect different approximate symmetries and
effective degrees of freedom. In the hadron gas
the degrees of freedom are well separated hadrons and chiral symmetry is spontaneously broken. In the stringy fluid chiral symmetry is restored and approximate chiral spin symmetry
emerges, which is a symmetry of confining interaction. One could view this medium as
a densely packed system of the color-singlet clusters. The quark interchanges between clusters required by Pauli principle should be
significant. Still it is a system with confinement. The QGP matter at high temperatures
consists of decofined quark and gluon  quasiparticles. 

The large $N_c$ limit of QCD with massless
quarks should clarify issues 
since in this case QCD with light quarks is manifestly center symmetric. This allows one to
define unambiguously possible phases  with confinement or deconfinement and with spontaneously broken or restored chiral symmetry. Standard large $N_c$ analysis
\cite{CG} suggests
that three regimes at $N_c=3$ and small but
nonzero quark masses might become distinct phases separated by phase transitions with
different scaling of main thermodynamical
quantities: energy density $\epsilon$, pressure  $P$ and entropy density $s$:

\begin{equation}
    \epsilon_{\rm HG} \sim N_c^0 \; \; , \; \; P_{\rm HG} \sim N_c^0 \; , \; s_{\rm HG} \sim N_c^0 \; , \label{Eq:had scale}
\end{equation}
 
\begin{equation}
    \epsilon_{\rm str} \sim N_c^1 \; \; , \; \; P_{\rm str} \sim N_c^1 \;  ,  \; s_{\rm str} \sim N_c^1 \;  ,
     \label{Eq:IntScal}
\end{equation}

\begin{equation}
    \epsilon_{\rm QGP} \sim N_c^2 \; \; , \; \; P_{\rm QGP} \sim N_c^2 \;  ,  \; s_{\rm QGP} \sim N_c^2 \;  .
     \label{Eq:IntScal}
     \end{equation}

A peculiar feature of the intermediate phase is that it should contain
a gas of noninteracting glueballs, as in the hadron gas, while the degrees
of freedom containing quarks are radically changed.
Lattice large $N_c$ studies could clarify this picture.

It is known that the deconfinement temperature in pure Yang-Mills theory
is practically $N_c$-independent \cite{LP}. At the same time confining
properties of QCD with light quarks are identical with those in Yang-Mills
at infinite $N_c$. Then one expects a deconfinement temperature in large
$N_c$ QCD with light quarks to be around $T_d \sim 300$ MeV. It is also found
on the lattice that at $T=0$ the quark condensate scales practically exactly as
$N_c^1$ and the $1/N_c$ corrections are negligibly small \cite{bon}. This strongly
suggests that the physics of chiral symmetry breaking in QCD at $N_c=3$ and at large $N_c$
is the same. Then one expects a chiral restoration phase transition in large $N_c$
QCD around the same temperature as at $N_c=3$. The latter temperature was established
to be around $T_{ch} \sim 130$ MeV \cite{Karsch}. If correct, one anticipates two different
phase transitions at large $N_c$, as was discussed above.

 \section{Outlook}
 The next important step to clarify the
 QCD phase diagram would be to perform
 lattice measurements of the chiral restoration temperature at large $N_c$. It is important to stress that the chiral and large $N_c$ limits do not commute and a proper sequence of limits should be taken. First the large $N_c$ limit and then the chiral limit. This sequence of limits
 significantly simplifies the numerical problem.
 As the first step the pure glue (quenched)
 configurations should be generated at large $N_c$. There are two strategies of doing this.
 The first strategy would be to employ
gluonic ensembles at large volumes
at reasonably large $N_c \sim 10-20$ as was done in Ref. \cite{LP}. Then the eigenvalue problem for the
Dirac operator should be solved to determine
the quark condensate via the Banks-Casher
relation. Upon increasing temperature the temperature
of the chiral restoration phase transition could be obtained. The alternative strategy would
be to follow the Eguchi-Kawai volume independence
to address the problem at very large $N_c$ but in a not  large volume, as was done in Ref. \cite{bon}. The results for chiral restoration temperatures should give in principle the same answer
in both approaches.


\end{document}